\documentstyle[aps,multicol,epsf,rotate,prl,psfig]{revtex}
\begin{document}
\title{A new upper limit on  the total neutrino mass from the 2dF Galaxy
Redshift Survey}
\author{\O. Elgar\o y$^1$, O. Lahav$^1$,
W. J. Percival$^2$,
J. A. Peacock$^2$,
D. S. Madgwick$^1$,
S. L. Bridle$^1$,
C. M. Baugh$^3$,
I. K. Baldry$^4$,
J. Bland-Hawthorn$^5$, T. Bridges$^5$, R. Cannon$^5$, S. Cole$^3$,
M. Colless$^6$, C. Collins$^7$,
W. Couch$^8$, G. Dalton$^9$, R. De Propris$^8$,
S. P. Driver$^{10}$, G. P. Efstathiou$^1$,
R. S. Ellis$^{11}$, C. S. Frenk$^3$, K. Glazebrook$^{5}$, C. Jackson$^6$,
I. Lewis$^9$,
S. Lumsden$^{12}$, S. Maddox$^{13}$,
P. Norberg$^3$,
B. A. Peterson$^6$, W. Sutherland$^2$, K. Taylor$^{11}$
}
\address{$^1$ Institute of Astronomy, University of Cambridge, Madingley Road,
Cambridge CB3 0HA, UK
}
\address{$^2$ Institute for Astronomy, University of Edinburgh,
Royal Observatory,
Blackford Hill, Edingburgh EH9 3HJ, UK}
\address{$^3$ Department of Physics, University of Durham, South Road,
Durham DH1 3LE, UK}
\address{$^{4}$ Department of Physics \& Astronomy, John Hopkins University,
Baltimore, MD 21218-2686, USA}
\address{$^5$ Anglo-Australian Observatory, P. O. Box 296, Epping, NSW 2121,
Australia}
\address{$^6$ Research School of Astronomy \& Astrophysics, The Australian
National University, Weston Creek, ACT 2611, Australia}
\address{$^7$ Astrophysics Research Institute, Liverpool
John Moores University, Twelve Quays House, Birkenhead, L14 1LD, UK}
\address{$^8$ Department of Astrophysics, University of New South Wales,
Sydney, NSW 2052, Australia}
\address{$^9$ Department of Physics, University of Oxford, Keble Road,
Oxford OX1 3RH, UK}
\address{$^{10}$ School of Physics and Astronomy, University of St. Andrews,
North Haugh, St. Andrews, Fife, KY6 9SS, UK}
\address{$^{11}$ Department of Astronomy, California Institute of Technology,
Pasadena, CA 91125, USA}
\address{$^{12}$ Department of Physics, University of Leeds, Woodhouse
Lane, Leeds, LS2 9JT, UK}
\address{$^{13}$ School of Physics \& Astronomy, University of Nottingham,
Nottingham NG7 2RD, UK}
\maketitle

\begin{abstract}

We constrain $f_{\nu} \equiv \Omega_{\nu}/\Omega_{\rm m}$, the fractional
contribution of neutrinos to the total mass density in the Universe, by
comparing the power spectrum of fluctuations derived from the 2dF
Galaxy Redshift Survey with power spectra for models with four components:
baryons, cold dark matter, massive neutrinos and a
cosmological constant.  Adding constraints from  independent
cosmological probes we find $f_{\nu} < 0.13$
(at 95\% confidence) for a prior of $0.1< \Omega_{\rm m} <0.5$, and 
assuming the scalar spectral index $n=1$.   
This translates to an upper limit on the total neutrino mass
$m_{\nu,\rm tot}< 1.8\;{\rm eV}$ for `concordance' values of $\Omega_{\rm m}$
and the Hubble constant.
Very similar results are obtained with a prior on $\Omega_{\rm m}$ from 
Type Ia supernovae surveys, and with marginalization over $n$.  

\end{abstract}

\begin{multicols}{2}

Whether neutrinos are massive or not has been
an open question for a long time, but the recent data from atmospheric
and solar neutrino experiments
\cite{superk,macro,homestake,gallex,sage,gno,sno}
are most naturally interpreted in
terms of neutrino oscillations, which implies that not all neutrinos
are massless \cite{creminelli,fogli,bahcall}.
However, since the oscillation probability
depends on the mass-squared differences,
and not on the absolute
masses, the oscillation experiments cannot provide absolute masses
for neutrinos.
The mass scale can in principle be obtained from e.g. the energy spectrum in
the beta-decay of $^{3}{\rm H}$ \cite{mainz}, or from neutrinoless double
beta decay \cite{klapdor1,klapdor2}.
At the present, cosmological data
\cite{fukugita,croft,gawiser}
provide stronger constraints on the total
neutrino mass than particle physics experiments.
Since neutrinos with masses of the order of a few tenths of an electron volt
(eV) can
have a significant effect on the formation of large-scale structures
in the Universe, observations of the distribution of galaxies can provide
us with an upper bound on the density of massive neutrinos.
We will in this paper use data from the 2dF Galaxy Redshift Survey (2dFGRS),
which is the largest existing redshift survey
\cite{colless,percival}, to obtain an upper bound on the sum of the
neutrino masses.

Massive neutrinos make up part of the dark matter
in the Universe.  In the cosmological model favoured by data
on large-scale structure and the observed fluctuations in the
Cosmic Microwave Background (CMB) \cite{efstathiou,wang}, the Universe
is flat, and the contributions to the
mass-energy density in units of the critical density are
$\Omega_\Lambda \approx 0.7$  from vacuum energy or  a `quintessence' field,
and $\Omega_{\rm m} \approx  0.3$ from matter.  Baryons
make up only $f_{\rm b} \equiv \Omega_{\rm b}/\Omega_{\rm m} \approx 0.15$
of the matter contribution \cite{efstathiou,wang},
most of the remaining being in the form of cold dark matter (CDM), the
exact nature of which is still unknown.  `Cold' in this context
means that the particles were moving at non-relativistic speeds
when they dropped out of thermal equilibrium.  Particles drop out
of equilibrium roughly when their interaction rate falls below the expansion
rate of the Universe.  For neutrinos with masses in the eV range
this happened when they were still relativistic, and so they will
be a `hot' component of the dark matter (HDM).  This has
implications for large-scale structure, since the neutrinos can
free-stream over large distances and erase small-scale structures
(see e.g. \cite{primack} for an overview).
As a result, mass fluctuations are suppressed at comoving wavenumbers greater than
$k_{\rm nr}=0.026
(m_{\nu}/1\;{\rm eV})^{1/2}\Omega_{\rm m}^{1/2}
\;h\,{\rm Mpc}^{-1}$ \cite{doro},
where $m_{\nu}$ is the neutrino mass of one flavour.
The neutrino contribution to the total mass-energy density, $\Omega_\nu$,
in units of the critical density needed to close the Universe,
is given by
\begin{equation}
\Omega_\nu h^2 = \frac{m_{\nu,\rm tot}}{94 \;{\rm eV}},
\label{eq:onurel}
\end{equation}
where $m_{\nu, \rm {tot}}$ is the sum of the neutrino
mass matrix eigenvalues,
and
the Hubble parameter $H_0$ at the present epoch is given in terms of $h$ as
$H_0 = 100\,h\;{\rm km}\,{\rm s}^{-1}\,{\rm Mpc}^{-1}$.  Eq. (\ref{eq:onurel})
assumes that all three neutrino flavours drop out of equilibrium
at the same temperature, which is a reasonable approximation.
The suppression of the matter power spectrum
$P_{\rm m}(k)$ on small scales is  given approximately by (see e.g. \cite{tegmark})
\begin{equation}
\frac{\Delta P_{\rm m}}{P_{\rm m}}   \approx -8 f_{\nu},
\label{eq:suppr}
\end{equation}
where
$f_{\nu} \equiv \Omega_\nu/\Omega_{\rm m}$.
Therefore even a neutrino mass as small as $0.1\;{\rm eV}$ gives a reduction
in power of 5-15 \%.

The 2dFGRS has now measured over 220 000
galaxy redshifts, with a median redshift of $z_m\approx 0.11$, and is the
largest existing galaxy redshift survey \cite{colless}.
A sample of this size allows large-scale structure statistics to be measured
with very small random errors.
An initial estimate of the convolved, redshift-space power
spectrum of the 2dFGRS has been determined \cite{percival} for
a sample of 160 000 redshifts.
On scales $0.02 < k < 0.15 \;h\,{\rm Mpc}^{-1}$, the data are
robust and the shape of the power spectrum is not affected by
redshift-space or non-linear effects, though the amplitude is increased
by redshift-space distortions.
These data and their associated covariance matrix form the basis for our analysis.

For each model, we calculate its linear-theory matter power spectrum,
and for the 2dFGRS power spectrum data it is sufficiently
accurate to use the fitting formulae derived in \cite{eisenstein}.
The relation between the measured galaxy
power spectrum and the calculated matter power spectrum is given
by the so-called bias parameter $b^2 \equiv P_g(k)/P_m(k)$.
By definition, $b$ is in principle a function of scale, and several  
biasing scenarios have been proposed \cite{dekel}.  This issue is of some importance 
for our analysis.  For example, if the galaxy distribution is more biased on small 
scales than on large scales, a non-zero best-fit value for $f_\nu$ may be obtained.  
However, on the scales we consider there are theoretical 
reasons to expect that $b$ should tend to a constant 
\cite{benson}.  For the 2dFGRS, a recent analysis \cite{verde} 
looking for deviations from 
linear biasing found no evidence for it.  We will therefore 
in the following assume that the biasing is scale-independent.  
In particular 
two independent  analyses \cite{lahav,verde} suggest that
that the data are consistent with $b \approx 1$
on large scales.  
We choose to avoid the complications
in the normalization of the power spectra caused by
redshift-space distortions and the actual value of the bias parameter
by leaving the amplitude of the power spectrum,  from here on 
denoted by $A$, as a free parameter, and then marginalize over it. 

\begin{figure}
\begin{center}
{\centering
\mbox
{\psfig{figure=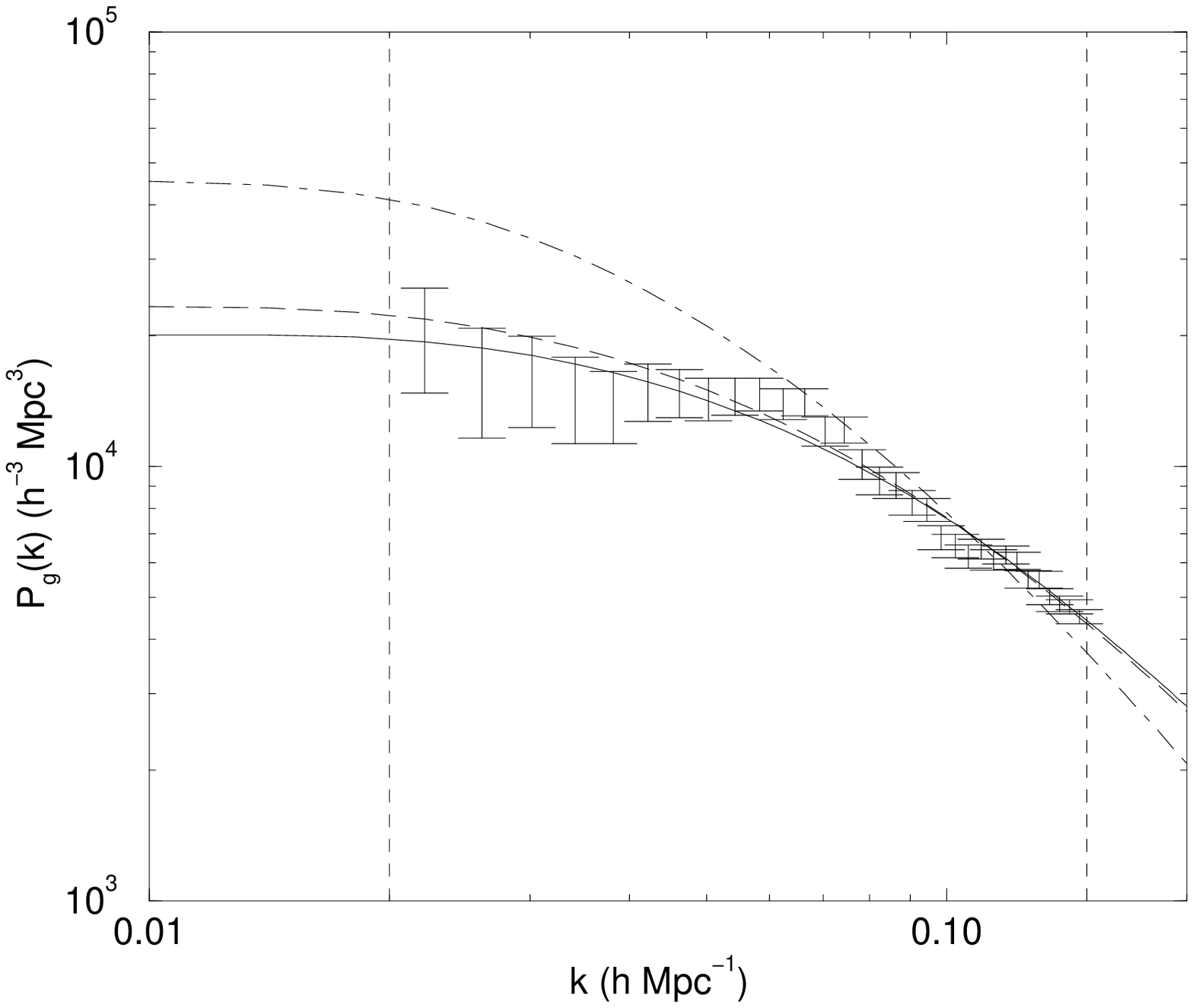,height=6cm,width=6cm}}
}
\caption{Power spectra for $\Omega_\nu = 0$ (solid line),
$\Omega_\nu=0.01$ (dashed line), and $\Omega_\nu=0.05$ (dot-dashed line)
with amplitudes fitted to the 2dFGRS power spectrum data (vertical bars) in
redshift space.  We have fixed
$\Omega_{\rm m}=0.3$, $\Omega_\Lambda=0.7$, $h=0.7$, $\Omega_{\rm b}h^2=0.02$.
The vertical dashed lines limit the range in $k$ used in the fits.}
\label{fig:fig1}
\end{center}
\end{figure}

We shall consider here a model with four components: baryons, cold
dark matter, massive neutrinos (hot dark matter) and a cosmological constant.
As an illustration, we show in Fig. \ref{fig:fig1} the power spectra
for  $\Omega_\nu = 0$, $0.01$, and $0.05$ (all other
parameters are fixed at their `concordance model' values given
in the figure caption),
after they have been convolved with the 2dFGRS window function, and their amplitudes
fitted to the 2dFGRS power spectrum data.  For the 32 data points, the
$\Omega_\nu=0$-model had $\chi^2=32.9$, $\Omega_\nu=0.01$ gives $\chi^2=33.4$,
whereas the model with
$\Omega_\nu=0.05$ provides a poor fit to the data with $\chi^2=92.2$.

Clearly, the inference of the neutrino mass depends on our
assumptions (`priors') on the other parameters.  We therefore add
constraints from other independent cosmological probes.  The Hubble
parameter has been determined by the HST Hubble key project to be
$h=0.70\pm 0.07$ \cite{freedman}, and Big Bang Nucleosynthesis gives a
constraint $\Omega_{\rm b}h^2 =0.020\pm 0.002$ on the baryon
density\cite{burles}.  For these parameters, we adopt Gaussian priors
with the standard deviations given above.  

Perhaps the least known prior is the total matter density $\Omega_{\rm m}$.
The position of the first peak in the CMB power spectrum gives a strong indication that the
Universe is spatially flat, i.e.  $\Omega_{\rm m} + \Omega_\Lambda = 1$
\cite{efstathiou,wang}.  The CMB peak positions are not sensitive to
neutrino masses, because the neutrinos were non-relativistic at
recombination, and hence indistinguishable from cold dark matter.
Although the shape of the power spectrum is independent of  curvature, 
the curvature does affect the choice of priors on $\Omega_{\rm m}$, and 
we choose to consider flat models only.   
When the constraint of a flat universe is combined with
 surveys of high redshift
Type Ia supernovae \cite{perlmutter,riess},
one finds $\Omega_{\rm m} =0.28\pm 0.14$.
However, studies of the mass-to-light
ratio of galaxy clusters find values of $\Omega_{\rm m}$ as low as 0.15
\cite{nbahcall}, whereas  cluster abundances give a range of values
$\Omega_{\rm m} \approx 0.3-0.9$
\cite{pierpaoli,seljak,reiprich,viana,blanchard,eke}.
Another measurement of $\Omega_{\rm m} \approx 0.25$
which is independent of the power spectrum of
mass fluctuations and the nature of dark matter
has been derived
from the baryon mass fraction in clusters of galaxies,
coupled with priors on  $\Omega_{\rm b} h^2$ and $h$
\cite{white,pirin}.
We will therefore use two different priors on $\Omega_{\rm m}$.
The first is a Gaussian centered at $\Omega_{\rm m}=0.28$
with standard deviation $0.14$, motivated by \cite{perlmutter}.
As an alternative, we use a uniform (`top hat') prior in the
range $0.1 <\Omega_{\rm m} < 0.5$.  Given that we use the
HST Key Project result \cite{freedman} for $h$, $\Omega_{\rm m} < 0.5$
is required to be consistent with the age of the Universe \cite{krauss} being
greater than 12 Gyr.

Finally, the CMB data \cite{wang,rubino,sievers} 
are consistent with the scalar spectral index of the primordial power
spectrum being $n=1$, but we also quote results
for $n=0.9$ and $n=1.1$, and for the case when we marginalize over 
$n$ with a Gaussian prior $n=1.0\pm 0.1$, motivated by the 
latest data from VSA and CBI \cite{rubino,sievers}.   

Results will be presented for the case of $N_\nu=3$ equal-mass
neutrinos, but the derived upper bound on the total neutrino mass is
only marginally different for $N_\nu=1$ or 2.  For each set of
parameters, we computed the theoretical matter power spectrum, and
obtained the $\chi ^2$ for the model given the 2dFGRS power spectrum.
We then calculated the joint
probability distribution function for $f_{\nu}$ and $\Gamma \equiv
\Omega_{\rm m} h$ (which represents the shape of the CDM power spectrum) by
marginalizing over $A, h$ and $f_{\rm b}$ weighted by the priors given
above.  For $A$ we used a uniform prior in the interval $0.5 < A <
10$, where $A=1$ corresponds to the normalization of the `concordance
model', discussed in \cite{lahav}.  Using instead a prior uniform in
$\log A$, or fixing $A$ at the best-fit value had virtually no effect
on the results.  We evaluated the likelihood on a grid with $0.1 <
\Omega_{\rm m} h < 0.5$, $0 \leq f_\nu < 0.3$, $0 < f_{\rm b} < 0.3 $,
$0.4 < h < 0.9$, and $0.5 < A < 10$.  In \cite{percival} it was found
that the 2dFGRS data alone allow a solution with a high baryon
density $f_{\rm b} = 0.4$, in addition to a low density-low baryon density
solution.  However, given the above priors, the solution with high
baryon density gets little weight and the fitting-formulae
in \cite{eisenstein} are sufficiently accurate for the measured BBN baryon density.

\begin{figure}
\begin{center}
{\centering
\mbox
{\psfig{figure=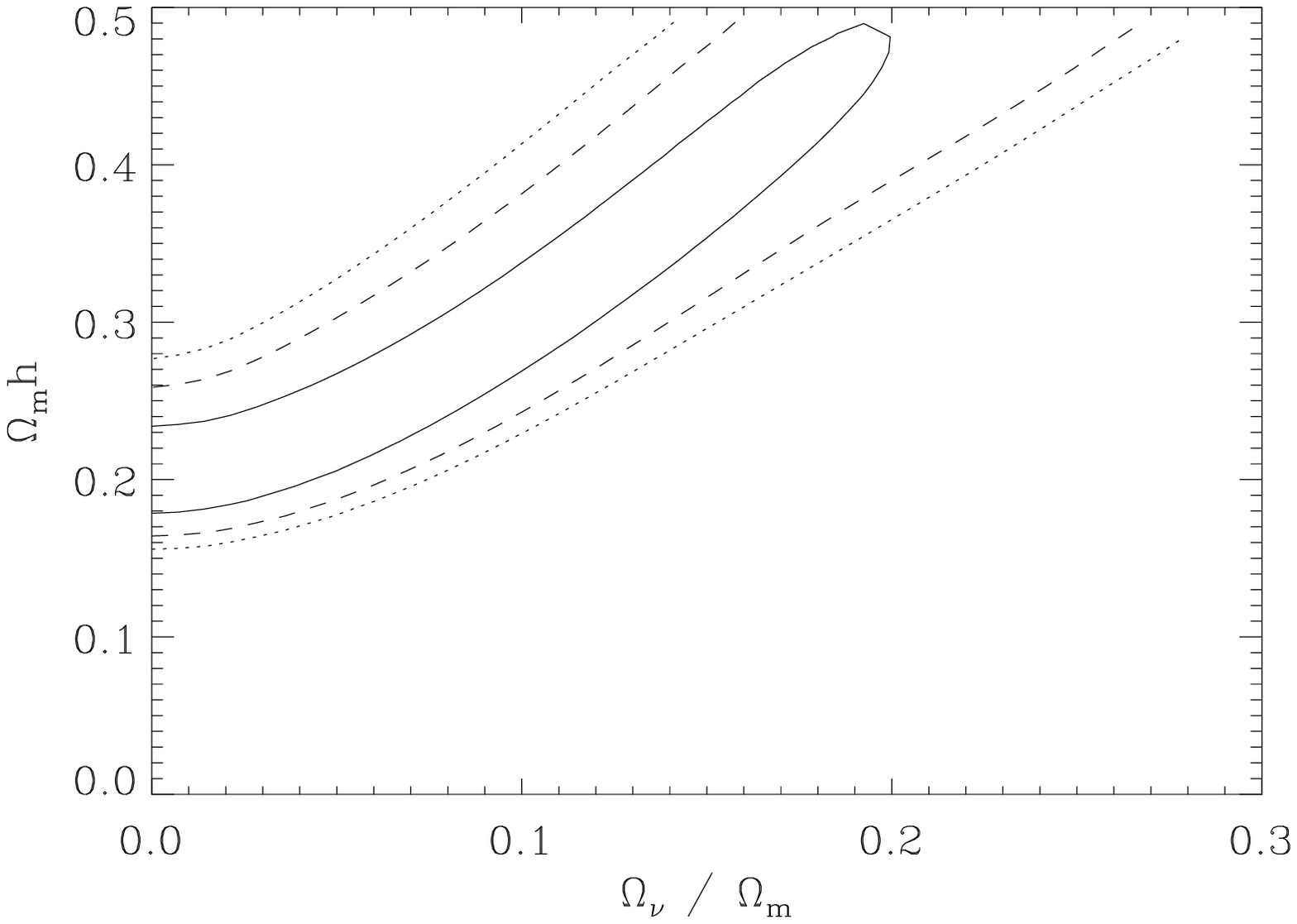,height=6cm,width=6cm}}
 \psfig{figure=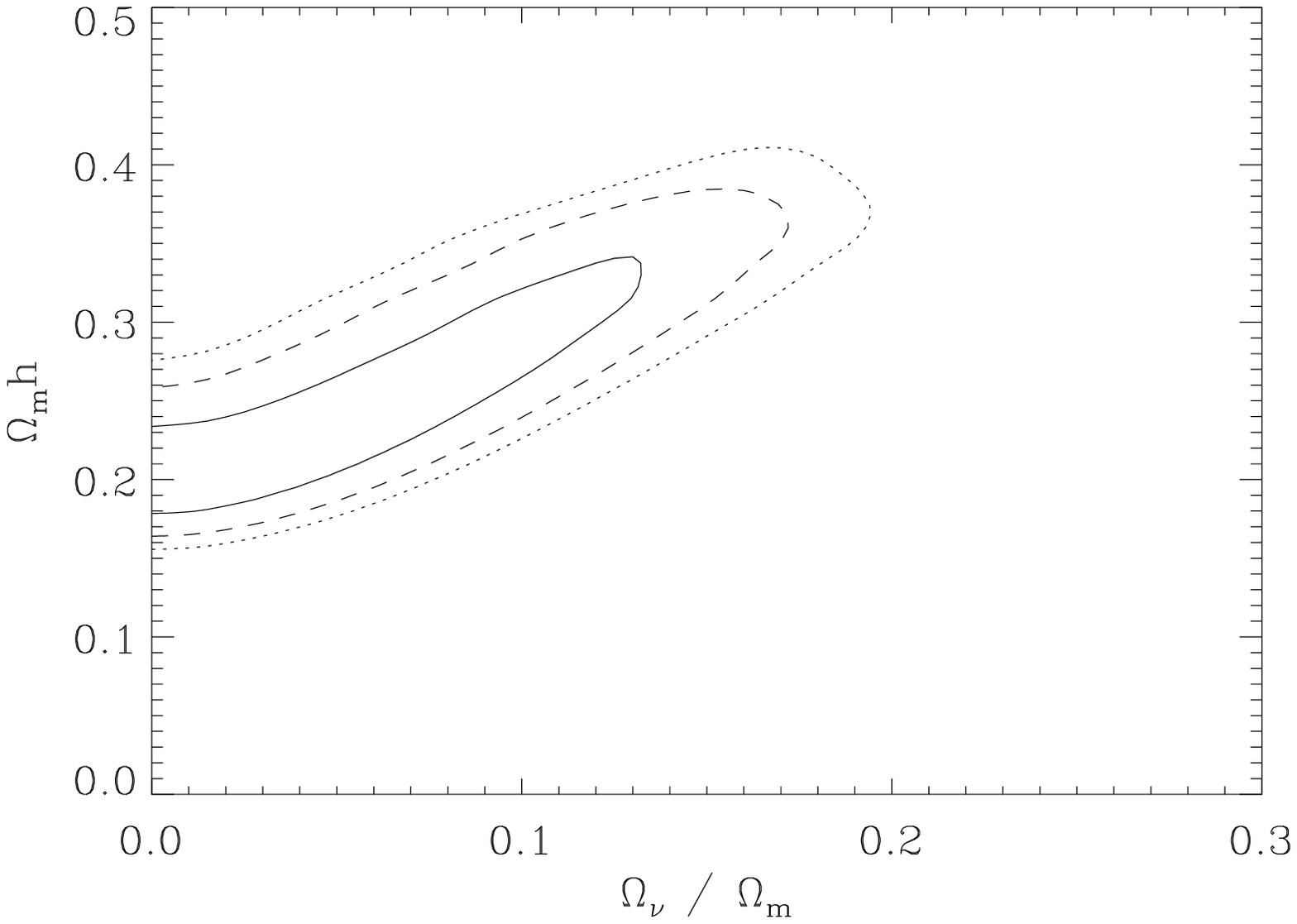,height=6cm,width=6cm}
}
\caption{68 (solid line), 95 (dashed line) and 99\% (dotted line)
confidence contours in the plane of
$f_\nu\equiv \Omega_{\nu}/\Omega_{\rm m}$
and $\Gamma\equiv\Omega_{\rm m} h$,
with marginalization over $h$ and $\Omega_{\rm b}h^2$ using Gaussian
priors, and over $A$ using a uniform prior in $0.5 < A < 10$.
The left panel shows the case of no prior on $\Omega_{\rm m}$,
and the right panel the case of a uniform `top hat' prior
on $\Omega_{\rm m}$ in $0.1<\Omega_{\rm m} < 0.5$.}
\label{fig:fig2}
\end{center}
\end{figure}

\begin{figure}
\begin{center}
{\centering
\mbox
{\psfig{figure=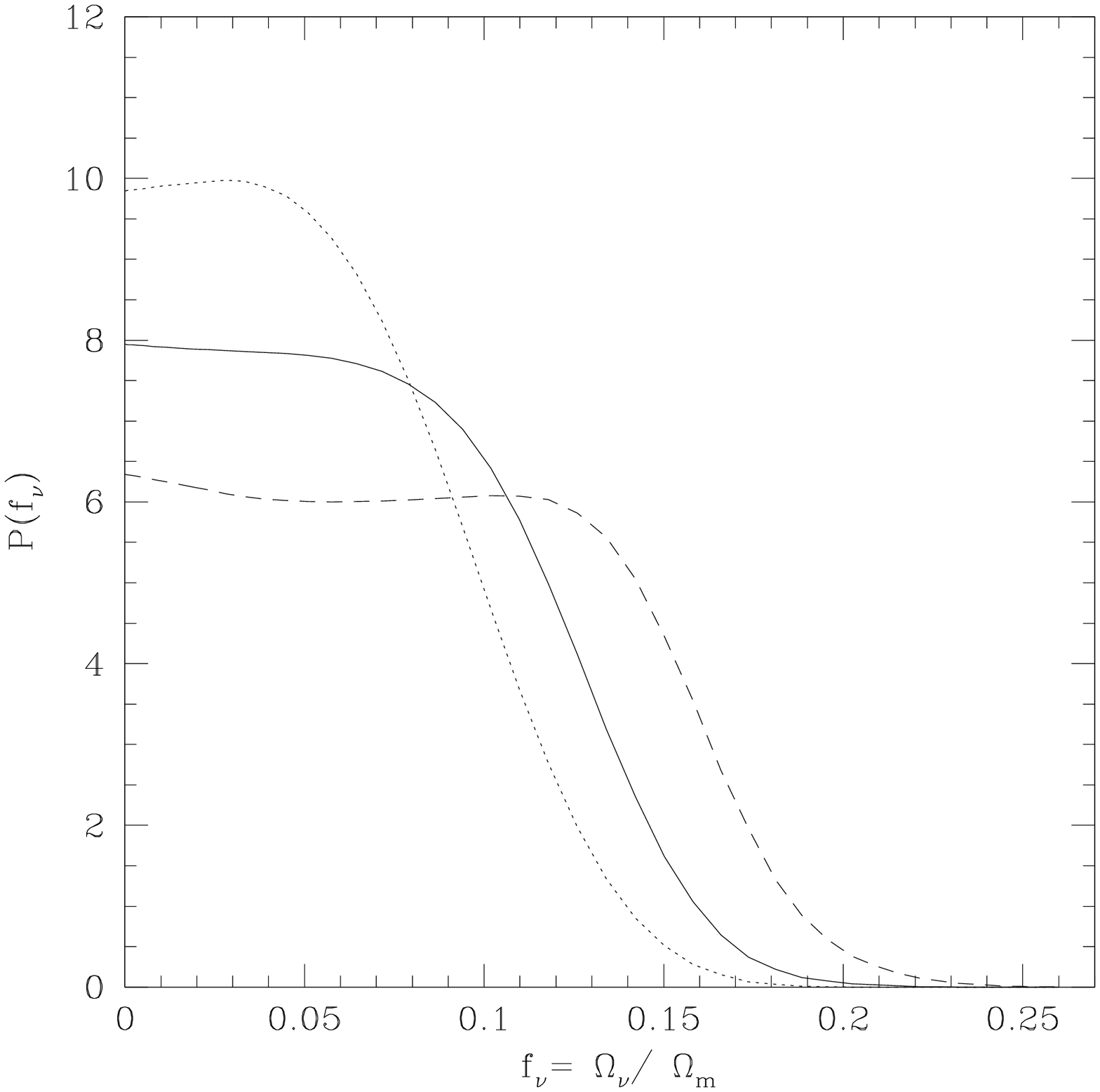,height=6cm,width=6cm}}
}
\caption{Probability distributions, normalized so that the area
under each curve is equal to one, for $f_\nu$
with marginalization over the other parameters, as explained in the
text, for $N_\nu =3$ massive neutrinos and $n=0.9$ (dotted line),
$1.0$ (solid line), and $1.1$ (dashed line).}
\label{fig:fig3}
\end{center}
\end{figure}

The results are shown in Fig. \ref{fig:fig2} for the cases of no prior
on $\Omega_{\rm m}$ (left panel) and with the uniform prior
$0.1 < \Omega_{\rm m} < 0.5$ (right panel).
Marginalizing the distributions in the right panel of Fig. \ref{fig:fig2}
over $\Omega_{\rm m} h$, we get the one-dimensional distribution for
$f_\nu$ given by the solid line in Fig. \ref{fig:fig3}, and an
upper limit $f_\nu < 0.13$ at 95\% confidence.  
For comparison, marginalizing without any priors, the limit becomes 
$f_\nu < 0.24$.  Adding just a prior on $\Omega_{\rm m}$, we find 
$f_\nu < 0.15$, so this is clearly the most important prior.  
Marginalizing with just a prior on $h$ or on $\Omega_{\rm b}h^2$, 
the 95 \% confidence limit becomes $f_\nu < 0.20$.      
As a further test of the stability of our analysis, we used 
the full set of priors, but only the power spectrum data at 
scales $k < 0.1\;h\,{\rm Mpc}^{-1}$. In this case 
the limit increases to $f_\nu < 0.20$. 

There is a further degeneracy of $f_\nu$ with the scalar spectral index $n$,
since increasing $n$ increases power on small scales and leaves
more room for suppression by the massive neutrinos.
Also shown in Fig. \ref{fig:fig3}
are the distributions for the cases $n=0.9$ (dotted line) and $n=1.1$
(dashed line).
With $n=1.1$, the 95\% confidence limit on $f_\nu$
increases slightly to 0.16.  The results are summarized
in Fig. \ref{fig:fig3} and in Table \ref{tab:tab1}.
Also included in the this table are the results obtained using
the Type Ia supernova prior, and it is seen that the results for
the two different choices are almost identical.  
Running a grid of models with $n$ as an added parameter, and marginalizing 
with a prior $n=1.0\pm0.1$, consistent with the CMB data \cite{rubino,sievers}, 
we find $f_\nu < 0.16$ at 95 \% confidence.  

To summarize, we have analyzed the shape of the 2dFGRS power spectrum to obtain an upper
bound on the fractional contribution of massive neutrinos to the
total mass density, $f_\nu$, and found an upper limit $f_\nu < 0.13$
at 95\% confidence for $0.1< \Omega_{\rm m} < 0.5$ and the scalar
spectral index $n=1$.  This translates into a constraint on the
sum of the neutrino mass matrix eigenvalues $m_{\nu,\rm tot}< 1.8\;{\rm eV}$
for $\Omega_{\rm m} h^2 = 0.15$.
With marginalization over $n$ with a prior $n=1.0\pm0.1$, 
the limit becomes $m_{\nu,\rm tot} < 2.2\;{\rm eV}$.
Previous cosmological bounds on neutrino masses come from data on
galaxy cluster abundances \cite{fukugita,lukash}, the Lyman $\alpha$ forest
\cite{croft}, and compilations of data including the CMB,
peculiar velocities, and large-scale structure \cite{gawiser}.
They give upper bounds on the total neutrino mass in the range 3--6 eV.
Note that the fluctuation amplitude derived from the cluster abundances
is still under some debate \cite{pierpaoli,seljak,reiprich,viana}.  
The most recent limit is that of
\cite{wang}, from a combined analysis of CMB and large-scale
structure data an upper limit $m_{\nu,\rm tot} < 4.2\;{\rm eV}$ was found.
Our bounds, summarized in Table \ref{tab:tab1}, are stronger, largely
because of the small statistical errors in the 2dFGRS power spectrum,
although the priors on the other parameters, in particular on
$\Omega_{\rm m}$, are also important.  Note also that all
of these results are stronger than current constraints from
particle physics.
As they stand, the controversial results of \cite{klapdor2}
imply a nearly degenerate neutrino mass matrix
and a bound on neutrino mass matrix eigenvalue
$0.1 < m_\nu < 1-20\;{\rm eV}$ \cite{feruglio,xing,barger}, where the 
upper limit is somewhat model-dependent.  Our results are consistent with this range.
If the largest neutrino mass is in fact of order a tenth of an eV,
it should be possible to measure its value using a combination of cosmological
data, combining 2dFGRS, SDSS, MAP and/or Planck.

\O E acknowledges support from the Research Council of Norway
through a postdoctoral fellowship.

\end{multicols}
\begin{table}
\begin{tabular}{lllll}
\multicolumn{1}{c}{ } &
\multicolumn{2}{c}{$\Omega_{\rm m}=0.28\pm 0.14$ (SNIa)} &
\multicolumn{2}{c}{$0.1<\Omega_{\rm m}<0.5$} \\
\multicolumn{1}{c}{$n$} &
\multicolumn{1}{c}{$f_\nu$} &
\multicolumn{1}{l}{$m_{\nu,\rm tot}$ (eV)} &
\multicolumn{1}{c}{$f_\nu$} &
\multicolumn{1}{c}{$m_{\nu,\rm tot}$ (eV)}\\ \hline
0.9  &  0.12  & 1.5 & 0.11 & 1.5 \\
1.0  &  0.14  & 1.8 & 0.13 & 1.8 \\
1.1  &  0.16  & 2.1 & 0.16 & 2.2
\end{tabular}
\caption{Summary of 95\% confidence upper bounds on
$f_\nu$ with our two chosen priors on $\Omega_{\rm m}$.
The conversion of $f_\nu$ to $m_{\nu,\rm tot}$ is for $h=0.7$ and
the central values of $\Omega_{\rm m}=0.28$ (SNIa case)
and  $\Omega_{\rm m}=0.30$ (uniform prior case).}
\label{tab:tab1}
\end{table}

\begin{references}
\bibitem{superk} S. Fukuda {\it et al.}, Phys. Rev. Lett. {\bf 85} (2000) 3999.
\bibitem{macro} M. Ambrosio {\it et al.}, Phys. Lett. B {\bf 517} (2001) 59.
\bibitem{homestake} B. T. Cleveland {\it et al.}, Ap. J. {\bf 496} (1998) 505.
\bibitem{gallex} The Gallex collaboration, Phys. Lett. B {\bf 447} (1999) 127.
\bibitem{sage} J. N. Abdurashitov {\it et al.}, Phys. Rev. C {\bf 60} (1999) 055801.
\bibitem{gno} M. Altmann {\it et al.}, Phys. Lett. B {\bf 490} (2000) 16.
\bibitem{sno} Q. R. Ahmad {\it et al.}, Phys. Rev. Lett. {\bf 87} (2001) 71301.
\bibitem{creminelli} P. Creminelli, G. Signorelli, and A. Strumia,
JHEP {\bf 0105} (2001) 52.
\bibitem{fogli} G. L. Fogli, E. Lisi, D. Montanino, and A. Palazzo,
Phys. Rev. D {\bf 64} (2001) 93007.
\bibitem{bahcall} J. N. Bahcall, M. C. Gonzalez-Garcia, and
C. Pena-Garay, JHEP {\bf 0108} (2001) 14.
\bibitem{mainz} J. Bonn {\it et al.}, Nucl. Phys. Proc. Suppl. {\bf 91} (2001) 280.
\bibitem{klapdor1} H. V. Klapdor-Kleingrothaus {\it et al.}, Eur. Phys. J. A
{\bf 12} (2001) 147.
\bibitem{klapdor2} H. V. Klapdor-Kleingrothaus {\it et al.}, Mod. Phys. Lett.
A {\bf 37} (2001) 2409.
\bibitem{fukugita} M. Fukugita, G-C. Liu, and N. Sugiyama,
Phys. Rev. Lett. {\bf 84} (2000) 1082.
\bibitem{croft} R. A. C. Croft, W. Hu, and R. Dav\'{e}, Phys. Rev. Lett.
{\bf 83} (1999) 1092.
\bibitem{gawiser} E. Gawiser and J. Silk, Science {\bf 280} (1998) 1405.
\bibitem{colless} M. Colless {\it et al.}, MNRAS {\bf 328} (2001) 1039.
\bibitem{percival} W. J. Percival {\it et al.}, MNRAS {\bf 327} (2001) 1297.
\bibitem{efstathiou} G. P. Efstathiou {\it et al}, MNRAS {\bf 330}(2002), L29.
\bibitem{wang} X. Wang, M. Tegmark, and M. Zaldarriaga, Phys. Rev. D {\bf 65} (2002) 123001.
\bibitem{primack} J. R. Primack, astro-ph/0112255. 
\bibitem{doro} A. G. Doroshkevich, Ya. B. Zel'dovich, R. A. Syunyaev, and
M. Yu. Khlopov, Soviet Astron. Lett. {\bf 6} (1980) 252.
\bibitem{tegmark} W. Hu, D. J. Eisenstein, and M. Tegmark, Phys. Rev. Lett.
{\bf 80} (1998) 5255.
\bibitem{eisenstein} D. J. Eisenstein and W. Hu, Ap. J. {\bf 511} (1999) 5.
\bibitem{dekel} A. Dekel and O. Lahav, Ap. J. {\bf 520} (1999) 24. 
\bibitem{benson} A. J. Benson, S. Cole, C. S. Frenk. C. M. Baugh, C. G. Lacey, 
MNRAS {\bf 311} (2000) 793. 
\bibitem{verde} L. Verde {\it et al.}, MNRAS, in press.
\bibitem{lahav} O. Lahav {\it et al.}, MNRAS, in press. 
\bibitem{freedman} W. L. Freedman {\it et al.} Ap. J. {\bf 553} (2001) 47.
\bibitem{burles} S. Burles, K. M. Nollett, and M. S. Turner, Phys. Rev. D
{\bf 63} (2001) 063512.
\bibitem{perlmutter} S. Perlmutter {\it et al.}, Ap. J. {\bf 517} (1999) 565.
\bibitem{riess} A. G. Riess {\it et al.} A.J. {\bf 117} (1999) 707.
\bibitem{nbahcall} N. A. Bahcall and J. M. Comerford, ApJL {\bf 565}(2002) L5.
\bibitem{pierpaoli} E. Pierpaoli, D. Scott, and M. White, MNRAS {\bf 325}
(2001) 77.
\bibitem{seljak} U. Seljak, astro-ph/0111362.
\bibitem{reiprich} T. H. Reiprich and H. Boehringer, Ap. J. {\bf 567} 716.
\bibitem{viana} P. T. P. Viana, R. C. Nichol, and A. R. Liddle, 
Ap. J. {\bf 569} (2002) L75.
\bibitem{blanchard} A. Blanchard, R. Sadat, J.G. Bartlett, and M. LeDour,
A\&A {\bf 362} (2000) 809.
\bibitem{eke} V. R. Eke, S. Cole, C. S. Frenk, and J. P. Henry,
MNRAS {\bf 298} (1998) 1145.
\bibitem{white} S. D. M. White, J. F. Navarro, A. E. Evrard, and C. S. Frenk,
Nature {\bf 366} (1993) 429.
\bibitem{pirin} P. Erdogdu, S. Ettori, and O. Lahav, astro-ph/0202357.
\bibitem{krauss}  L. M. Krauss and B. Chaboyer, astro-ph/0111597.
\bibitem{rubino} J. A. Rubi\~{n}o-Martin {\it et al.}, astro-ph/0205367.
\bibitem{sievers} J. L. Sievers {\it et al.}, astro-ph/0205387. 
\bibitem{lukash} N. A. Arhipova, T. Kahniashvili, and V. N. Lukash, A\&A {\bf 386} 
(2002) 775.
\bibitem{feruglio} F. Feruglio, A. Strumia, and F. Vissani, Nucl. Phys. B, in press.
\bibitem{xing} Z. Xing, Phys. Rev. D {\bf 65} (2002) 077302. 
\bibitem{barger} V. Barger, S. L. Glashow, D. Marfiata, and K. Whisnant, 
Phys. Lett. B {\bf 532} (2002) 15. 
\end{references}
\end{document}